\documentclass[a4paper]{jpconf}
\usepackage{graphicx}
\begin{document}
\title{Vortex-like state observed in ferromagnetic contacts}

\author{I. K. Yanson$^1$, Yu. G. Naidyuk$^1$, V. V. Fisun$^1$, O. P. Balkashin$^1$, L.~Yu. Triputen$^1$, A. Konovalenko$^2$
and V. Korenivski$^2$ }

\address{$^1$ B. Verkin Institute for Low Temperature Physics and Engineering,
National Academy of Sciences of Ukraine, 47 Lenin ave., 61103,
Kharkiv, Ukraine}

\address{$^2$ Nanostructure Physics, Royal Institute of Technology, SE-10691 Stockholm,
Sweden}

\ead{naidyuk@ilt.kharkov.ua}

\begin{abstract}
Point-contacts (PC) offer a simple way to create high current densities, 10$^9$ A/cm$^2$ and beyond, without substantial Joule heating. We have shown recently (Nano Letters, 7 (2007) 927) that conductivity of nanosized PCs between a normal and ferromagnetic metals exhibits bi-stable hysteretic states versus both bias current and external magnetic field -- the effect typical for spin-valve structures. Here we report that apart from the bi-stable state a third intermediate-resistance state is occasionally observed. We interpret this state as due to a spin-vortex in the PC, nucleated either by Oersted field of the bias current and/or by the circular geometry of PC.
The observed three-level-states in the PC conductivity testify that the interface spins are both weakly coupled to the spins in the bulk and have depressed exchange interaction within the surface layer.
\end{abstract}

Recently it was shown that a point contact (PC) between nonmagnetic (N) metals and a single nanometer thick ferromagnetic film (F-film) behaves as a spin-valve tri-layer structure F1-N-F2 \cite{Nanolett}. The explanation proposed is that the magnetization of a ferromagnetic surface layer (F1) placed at the contact N/F interface can be more freely rotated or even reversed compared to that of the interior F-film playing the role of the fixed F-layer (F2) in conventional spin-valve. The surface layer is atomically thin and plays a role of free layer (F1) in such surface spin-valve.
The resistance of PC is governed by the giant magnetoresistive effect in such a way that it contains two stable values: the larger one, corresponding to the anti-parallel (AP) magnetization F1 vs  F2, and the lower one when the magnetization in both F1 and F2 are parallel (P). The dynamical spin-valve behavior and static hysteresis in $dV/dI(V,H=0)$ or $dV/dI(H,V=0)$  characteristic of these contacts was firstly observed in \cite{Myers,Ji,Chen} and recently explained in \cite{Nanolett} introducing the surface spin-valve effect. The static hysteresis loop is due either to the transport current $I$ via the spin-transfer-torque (STT) effect \cite{Slonszewski} or to the external magnetic field $H$.

Surprisingly, while elaborating this effect we have found that besides the expected two resistance values of $dV/dI(V)=R(V)$ corresponding to P and AP states, a third resistance branch (state) was occasionally observed (Fig.\,1), which lays in between the former two.
These $R(V)$ characteristics appear to be very similar to what was observed in current perpendicular to plane (CPP) spin-valves geometry of small nanopillar rings with lateral dimensions  about 100\,nm \cite{Yang}, although the PC size (estimated from the resistance) is an order of magnitude smaller ($\sim$10\,nm). We assume that the third resistance value observed by us is also due to the vortex state (Vx) of magnetization structure in the atomically thin surface F-layer of a surface spin-valve \cite{Nanolett}, in spite of its extraordinary small size. Although a vortex state can be readily realized in circular geometry, an appearing of this configuration in mechanically prepared PC is more or less accidentally.

The dependence of $R(V,H)$ for mechanically produced PC between the Cu needle and the extended Co film on the applied voltage/current and external magnetic field was investigated. The layer structure in sequence of their deposition on a substrate consists of: i) the buffer layer of Cu (100\,nm), ii) Co layer (100\,nm), and, finally, iii) the capping Cu layer  (3\,nm) to prevent Co from oxidation. The contact is produced by gently touching a sharpened Cu needle on a free surface of this structure at low (liquid helium) temperature.
The external magnetic field was applied parallel to the film plane.
Typical contact resistances are of the order of 10 $\Omega$, which gives the lower limit of the contact diameter $d\sim$10\,nm by using the ballistic Sharvin formula \cite{book}.


\vspace{2cm}
\begin{figure}[h]
\begin{minipage}{18pc}
\includegraphics[width=18pc]{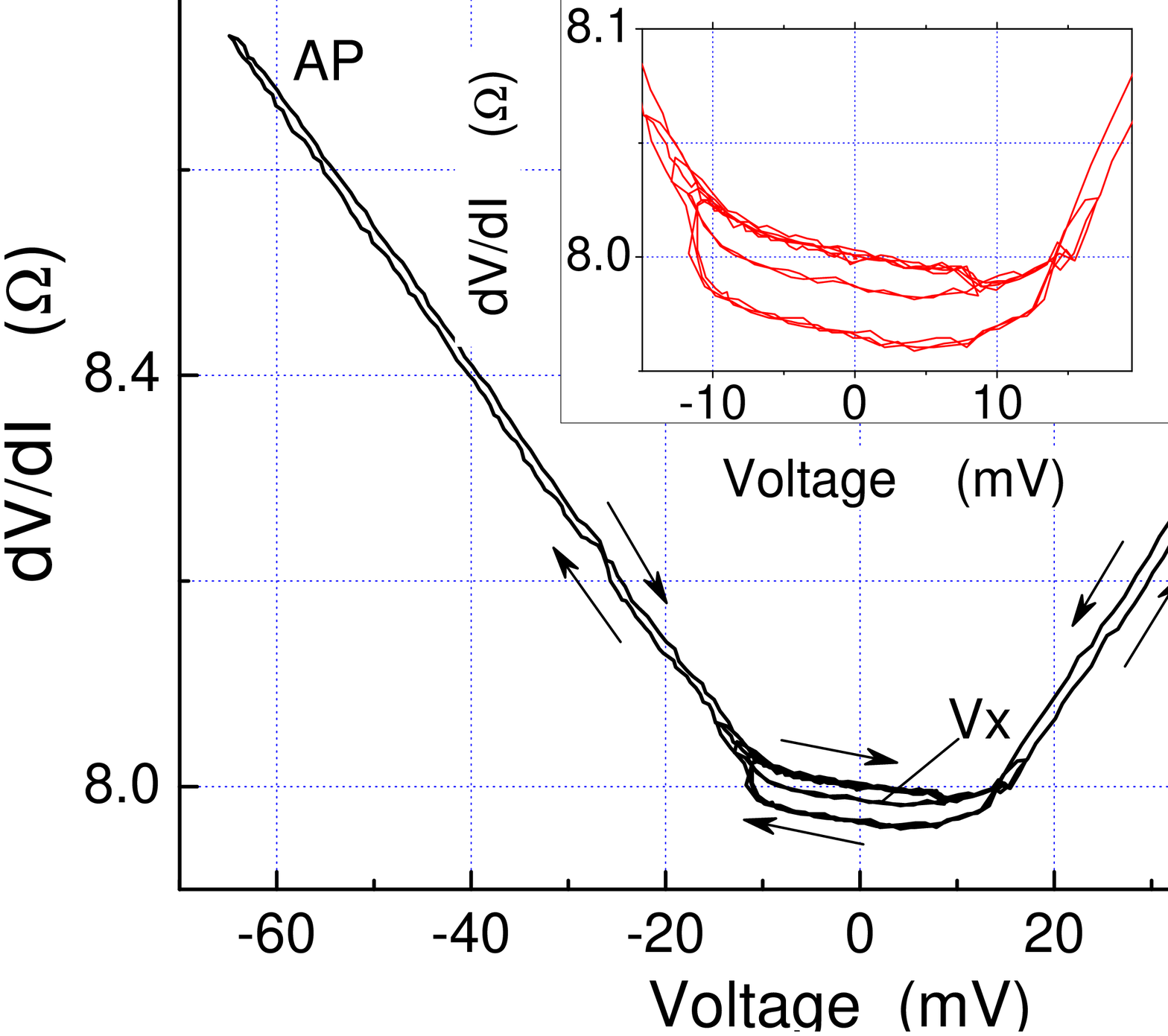}
\vspace{-3cm}
\caption{\label{f2}(Color online)
$dV/dI(V)$ showing the intermediate vortex (Vx) state inside the hysteretic spin-valve loop. The arrows show the sweep direction. The branches at large bias do not coincide showing the, so-called, lagging effect (see also text).  Inset: the magnified loop, where double repetition of AP-Vx cycles and triple repetition of AP-P cycles are shown (see also the Fig.\,2). }
\end{minipage}\hspace{2pc}
\begin{minipage}{18pc}
\includegraphics[width=18pc]{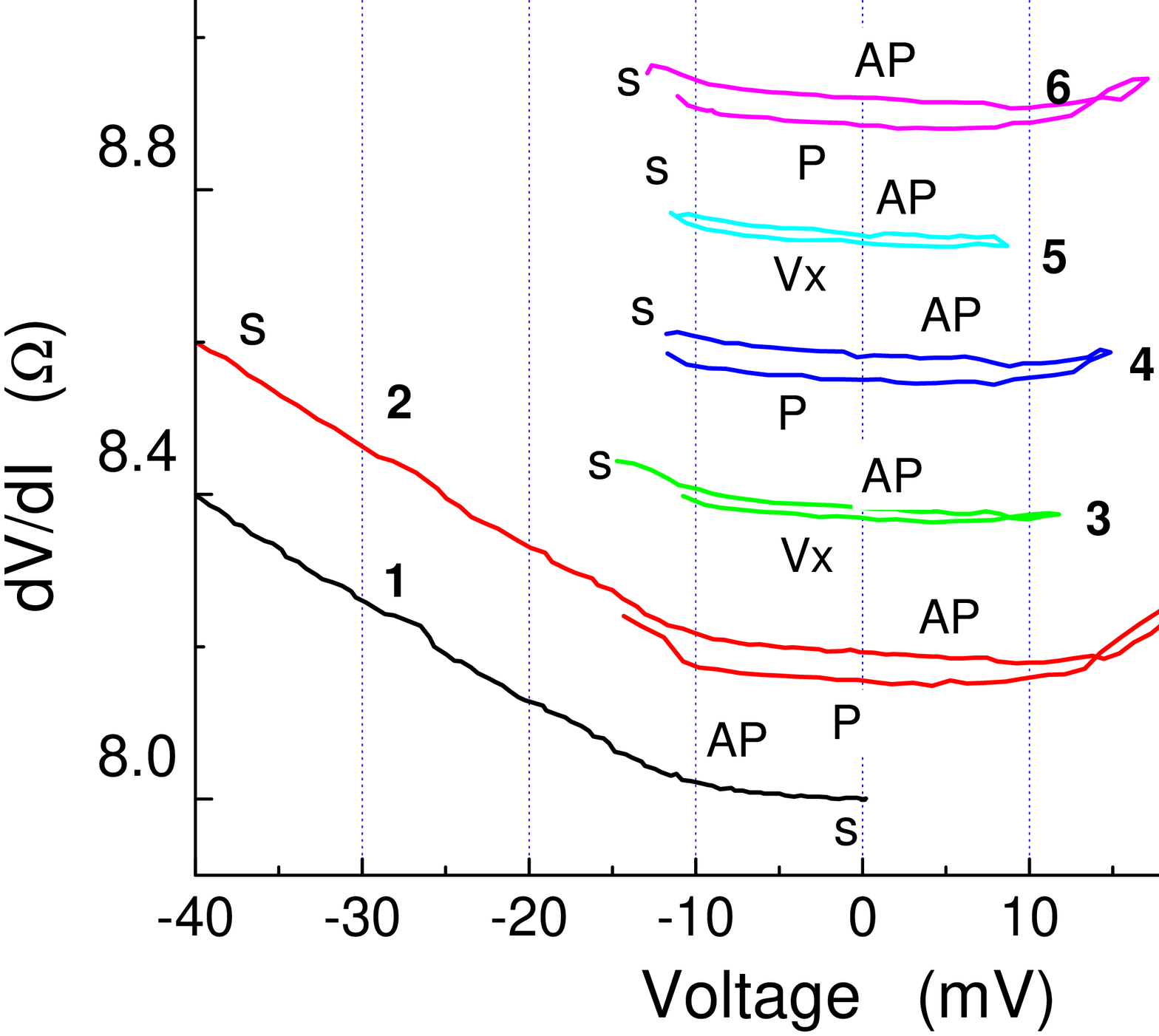}
\vspace{-3cm}
\caption{\label{f3}(Color online)
$dV/dI(V)$ from Fig.\,1 but separated on the successive cycles. The curves 2--6 are offset vertically for clarity. Letter $s$ notes the beginning of each cycle. The notations of branches AP, P, and Vx are given through their resistances in the $R(V)$ loop.
The sweeping direction is from negative to positive voltage and back, except for the curve 1.}
\end{minipage}
\end{figure}



Fig.\,1 shows the raw cumulative $R(V)$ curve of the PC with the mentioned intermediate state. It is clearly visible that the loop is divided in 3 repeatable branches whatever are the voltage/current limits of each particular cycles. For large limits ($V\simeq \pm$ 75\,mV) one can notice the lagging behavior of $R(V)$, which is not an experimental artifact but is an intrinsic characteristic of each particular junction likely due to its defect structure. That means the spins remember their previous directions for some time long enough. This feature is the property of specific magnetic structure at the interface and could be changing down to its complete absence (like, for example, shown in a number of Figs. to Ref.\,\cite{Nanolett}). The transition from the loop region in AP/P single branches occurs at any biases larger than approximately 10\,mV.

In Fig.\,2, the cumulative curve of Fig.\,1 is expanded on partial successive cycles (curves 1-6). Each cycle consists of forward direction half cycle (from -$V$ to +$V$ biases) and the return one. The initial point of the cycle is denoted by letter $s$ (start). The particular branch is also denoted by the magnetic state at 10\,mV bias as AP, P, or Vx, which are determined from the cumulated curve of Fig.\,1 according to three different branches of the loop pattern. One sees that regardless whatever is the voltage limits of each AP--P branch the height of the loop remains the same (curves 2, 4, 6). The same property is evident for the AP--Vx partial loop (curves 3, 5). It is quite important that in order to enter the Vx branch from AP branch one cannot exceed the bias $\sim$+15 mV, and has to return at bias $\sim$+10 mV.

Several successive magnetoresistance loops of the same PC are depicted in Fig.\,3. These are two meanders of different heights. The larger height of them corresponds to the height of AP-P loop in the $R(V$) curve of Fig.\,2. The external magnetic field can rotate both the upper atomically thin F-layer and the extended F-film. In any case, the CPP resistance is a measure concerning only the contact area. This is why the heights of hysteresis loops are the same in different physical measurements: namely, the $R(V)$ loop caused by STT force at $H$=0 and by external magnetic field force at $V$=0. In Fig.\,3 we again notice three stable levels of contact resistances. They are illustrated by the two constant vertical bars with arrows which are graphically superimposed on the measured $R(H,V=0)$ cycles. The ratio of larger to smaller bar heights equals approximately to the corresponding ratio from $R(V)$ curve.
\vspace{2cm}
\begin{figure}[h]
\begin{minipage}{18pc}
\includegraphics[width=19pc]{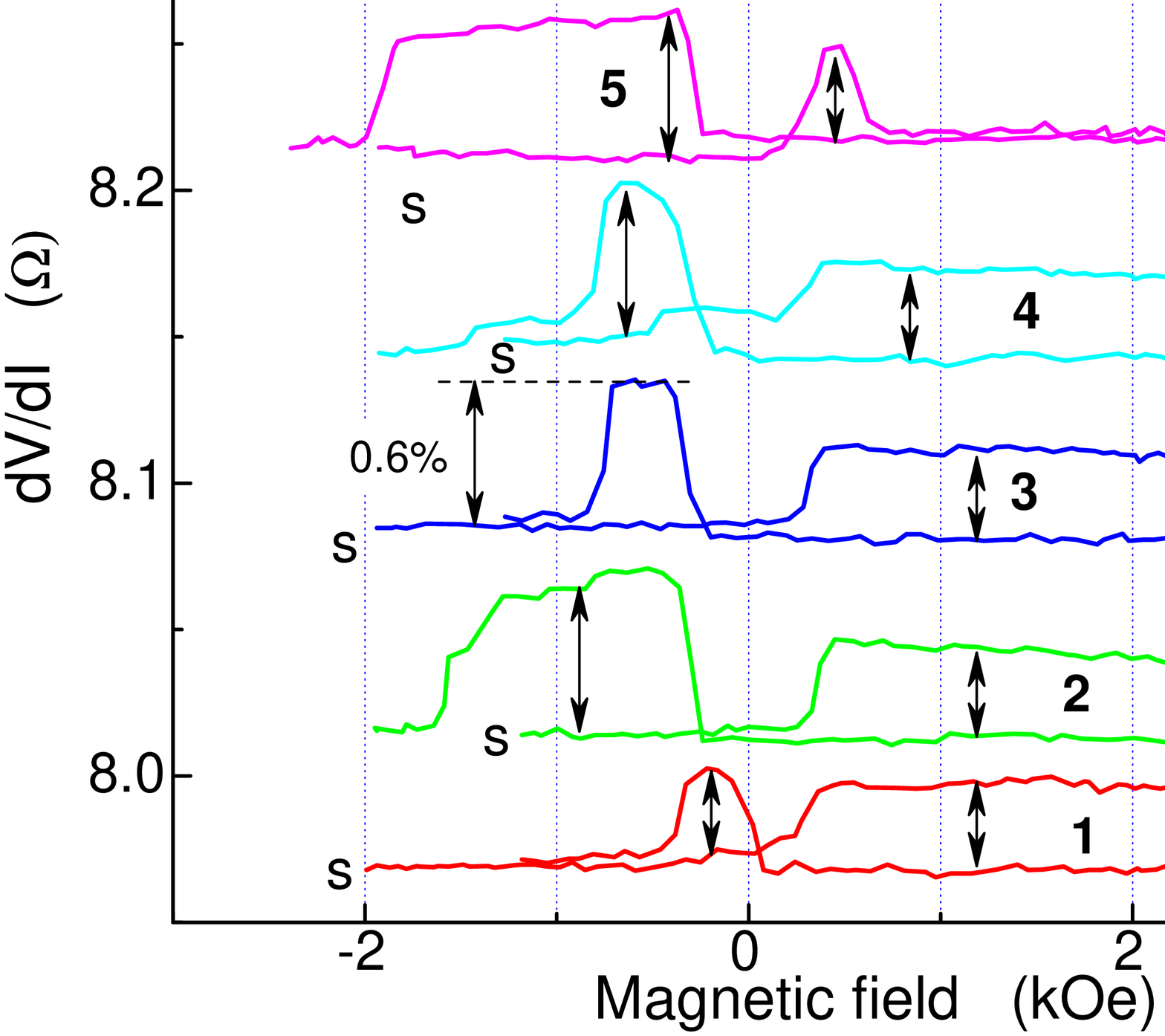}
\vspace{-3cm}
\caption{\label{f4}(Color online) 5 cycles of magnetoresistance dependence of the PC shown in Figs.\,1, 2. Curves 2 -- 5 are offsetting vertically for clarity. The two vertical arrows are graphically superimposed on the experimental curves showing the two stable resistance levels: between P and AP (longer arrow), and P and Vx (shorter arrow) states. $s$ labels the begin of each cycle. The reduced height of larger meander equals 0.6\%, the same as in Fig.\,1.}
\end{minipage}\hspace{2pc}%
\begin{minipage}{18pc}
\includegraphics[width=19pc]{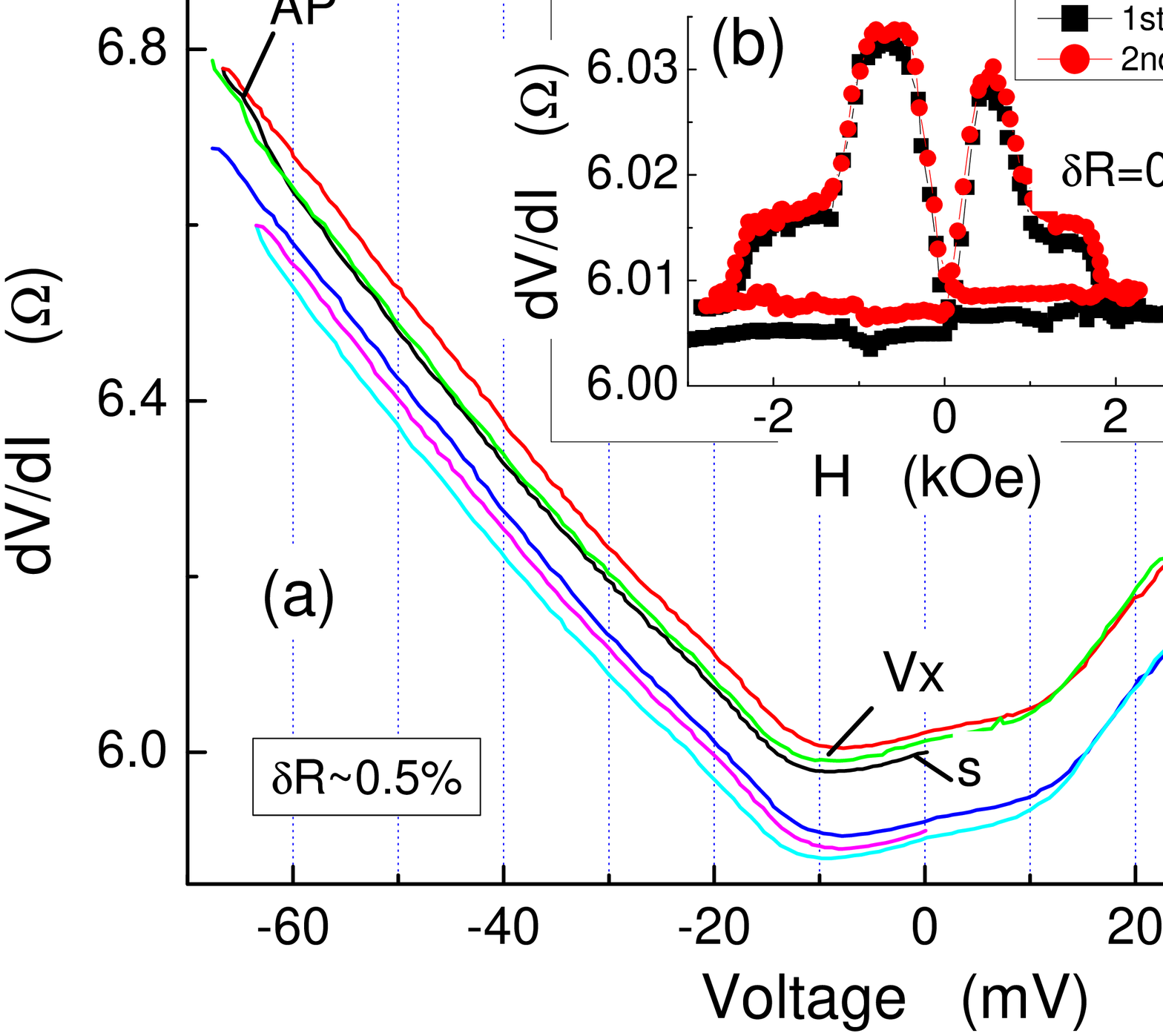}
\vspace{-3cm}
\caption{\label{f4}(Color online)
(a) Superimposed successive half-cycles of $R(V)$ dependences showing wide hysteresis loops. The half-cycles 4--6 are offset down by 0.1. Colors and magnetic states label each half-cycle. At ultimate large biases of negative and positive polarities the AP and P states are achieved, respectively.
(b) A couple of magnetoresistance cycles for the same PC. Data points are averaged over 5 adjacent experimental points.
}
\end{minipage}
\end{figure}

The next PC in Fig.\,4 demonstrates the three-branches $R(V)$ loop extending up to $V=-60$\,mV. At this relatively large bias ($\sim$--65\, mV), the magnetic configuration changes, leading to the smeared-step-like increase of the resistance from P or Vx to AP states, except for the third cycles (green color), where the Vx configuration changes to P one at an interval of approximately $-$(20--40)\,mV, which further goes up to the AP states at $V$=--65\,mV. Evidently, the pulling STT force is not large enough at smaller bias, in order to turn the system on the AP state immediately.
Very spectacular is that the magnetoresistance loops $R(H)$ shown in Fig.\,4b are very similar to what is observed for a nano-ring of quasi spin-valves described in \cite{Yang}. In the latter paper, the step structure of $R(H)$ within one loop is proven as an evidence of the existence of a vortex state in the free layer of conventional valve structure.
This is why we are inclined to interpret our additional branch as a vortex state labeling it by Vx, although we are quite aware that the observation of vortex state in so small sizes ($\sim$10\,nm) seems very challenging. We may argue that the exchange forces in the atomically thin surface layer (at least in a PC core) are noticeably smaller than in the bulk. That is why the lower limit imposed on the sizes for observation of the vortex state in the surface spin-valve can be essentially decreased. Also, one can anticipate that the CPP resistance of the vortex state lies in between the resistances of AP and P states, since in the vortex state one half of longitudinal (with respect to the bulk) component of interface spins are always directed along to bulk magnetization and the other half in the opposite direction.

	Following the consideration elaborated in \cite{Yang} there are two magnetic fields associated with current flowing trough the contact. One is the STT field which depends on the polarity of the current and for negative sign acts in our junctions in AP direction to the magnetization of the bulk of the film. Another is the so-called Oersted field which has a curling configuration and forces the spin pattern to get a vortex-like configuration. First one is proportional to the current density and for the given current magnitude the STT field is inversely proportional to the squared contact diameter. The second has vortex-like field which is inversely proportional to the contact diameter. Thus, for smaller diameter of the contact the STT field prevails at a given current and the final magnetic configuration at large current value is either single-domain AP or P state depending on the current polarity. For enabling vortex state the contact size should be neither too large nor too small. To estimate by order of magnitude the Oersted magnetic field we can use the expressions of \cite{Yang}. For free nano-ring of the spin-valve geometry and lateral contact size of about 100\,nm the Oersted field can amount 100\,Oe at a current of 10\,mA according to \cite{Yang}, while in our contact with lateral sizes $\sim$10\,nm it should be about 1\,kOe. This field is comparable with external magnetic field while changing the free layer orientation during the $R(H)$ dependences (see Fig.\,4b).
	Being established by the PC current, one of the three found stable states is conserved, when changing the current down to the remanent configuration with $V$=0 and $H$=0. This stability of a particular remanent state is due to the coercive force caused by the pinned spins depending on particular pattern of defects being specific for a given junction, until the contact jumps to more favorable energetic states with its own boundary current limits.

Thus, our main finding is the experimental observation of additional static magnetic states in PCs containing a single ferromagnetic film. Comparing with the previously found bistable states of the surface spin-valve \cite{Nanolett} we have found a third one. This state has a resistance lying approximately at the middle between well recognized resistance due to AP and P magnetization of atomically thin surface layer and the extended bulk of the ferromagnetic film \cite{Nanolett}. The latter is convincingly shown by comparing both $R(V)$ and $R(H)$ loops. It seems that the same is true concerning the new state which we tentatively named the vortex state, since its $R(V)$ and $R(H)$ characteristics strongly look like the vortex state in nano-ring spin-valve \cite{Yang}.

Finally, we have found the third stable remanent state in the surface spin-valve \cite{Nanolett}. This state seems to be very similar to the vortex state of spins as was found by comparison with nano-ring spin-valve \cite{Yang}. The spins are parallel to the film plane belonging to the atomically thin surface F-layer near the F/N interface.
We argue that these surface spin states, subject to a weakened exchange interaction, dominate the STT effects on the nanometer scale. The size of this vortex configuration is too small ($\sim$10\,nm) to be directly observed by any of existing technique, so far.

$Acknowledgments.$ The support of FP7 programm of EU under project STEELE $\#$225955 and "HAHO"-programm NAS of Ukraine under project $\#$02/09--H is acknowledged.

\end{document}